\documentclass{IEEEtran}

\usepackage[hidelinks]{hyperref}
\usepackage{caption}
\usepackage{indentfirst}

\usepackage{xcolor}
\usepackage{graphicx}
\usepackage{epstopdf}
\graphicspath{{fig/}}
\usepackage{subfigure}
\usepackage{float}
\usepackage{array}
\usepackage{url}
\usepackage{longtable}
\usepackage{rotating}
\usepackage{multirow}
\usepackage{amsmath}
\usepackage{amsfonts}
\usepackage{amssymb}
\usepackage{mathrsfs}
\usepackage{mdwlist}
\usepackage{booktabs}
\usepackage{threeparttable}
\usepackage{makecell}
\usepackage{diagbox}
\usepackage{pifont}
\usepackage{xcolor}
\usepackage{booktabs}
\usepackage{adjustbox}
 \usepackage{multirow}
\usepackage{enumitem}

\usepackage{pdflscape}
\usepackage{footnote}

\usepackage{bm}

\usepackage[numbers,sort&compress]{natbib}
\usepackage[linesnumbered,boxed,ruled]{algorithm2e}
\usepackage[hang,flushmargin]{footmisc}
\usepackage{lipsum}

\usepackage[T1]{fontenc}
\usepackage[utf8]{inputenc}

\usepackage{color,xcolor}
\definecolor{myred}{RGB}{217,46,127}
\definecolor{mygreen}{RGB}{67,127,127}

\title{FoleySpace: Vision-Aligned Binaural Spatial Audio Generation}

\author{
    {Lei Zhao, Rujin Chen, Chi Zhang, Xiao-Lei Zhang, \IEEEmembership{Senior Member, IEEE} and Xuelong Li, \IEEEmembership{Fellow, IEEE}}

    \thanks{Lei Zhao, Rujin Chen and Xiao-Lei Zhang are with the School of Marine Science and Technology, Northwestern Polytechnical University, Xi’an 710072, China, also with the Institute of Artificial Intelligence (TeleAI), China Telecom, P. R. China, and also with the Research and Development Institute of Northwestern Polytechnical University in Shenzhen, China (e-mail: zhao\_lei@mail.nwpu.edu.cn, chenrujin@mail.nwpu.edu.cn, xiaolei.zhang@nwpu.edu.cn).}
    \thanks{Chi Zhang and Xuelong Li are with the Institute of Artificial Intelligence (TeleAI), China Telecom, P. R. China (e-mail: zhangc120@chinatelecom.cn, xuelong\_li@ieee.org).}
    \thanks{The demo is available at \url{https://ralei96.github.io/FoleySpace/}.}
}

\begin{document}
\maketitle

\begin{abstract}
Recently, with the advancement of AIGC, deep learning-based video-to-audio (V2A) technology has garnered significant attention. However, existing research mostly focuses on mono audio generation that lacks spatial perception, while the exploration of binaural spatial audio generation technologies, which can provide a stronger sense of immersion, remains insufficient. To solve this problem, we propose FoleySpace, a framework for video-to-binaural audio generation that produces immersive and spatially consistent stereo sound guided by visual information. 
Specifically, we develop a sound source estimation method to determine the sound source 2D coordinates and depth in each video frame, and then employ a coordinate mapping mechanism to convert the 2D source positions into a 3D trajectory. This 3D trajectory, together with the monaural audio generated by a pre-trained V2A model, serves as a conditioning input for a diffusion model to generate spatially consistent binaural audio.
To support the generation of dynamic sound fields, we constructed a training dataset based on recorded Head-Related Impulse Responses that includes various sound source movement scenarios. 
Experimental results demonstrate that the proposed method outperforms existing approaches in spatial perception consistency, effectively enhancing the immersive quality of the audio-visual experience.

\end{abstract}


\section{Introduction} \label{sec:introduction}
Human visual and auditory systems rely on highly consistent audio-visual synchronization to perceive the external world. This multi-sensory integration not only enhances immersion but also improves contextual understanding and spatial awareness of video content. By simulating the content, timing, direction, distance, and movement trajectories of sound in the real world, audio-visual synchronization significantly enhances users' sense of presence and interaction quality in virtual environments. Therefore, automatically generating audio that is semantically, temporally, and spatially aligned with the video is of great importance for immersive extended reality (XR) applications, interactive entertainment systems, and dynamic media production.

Most early research on video-to-audio (V2A) generation primarily focuses on generating audio that is semantically consistent with the video content \cite{zhou2018visual, chen2020generating,iashin2021taming}.
\begin{figure}[t]
    \centering
    \includegraphics[width=0.90\columnwidth]{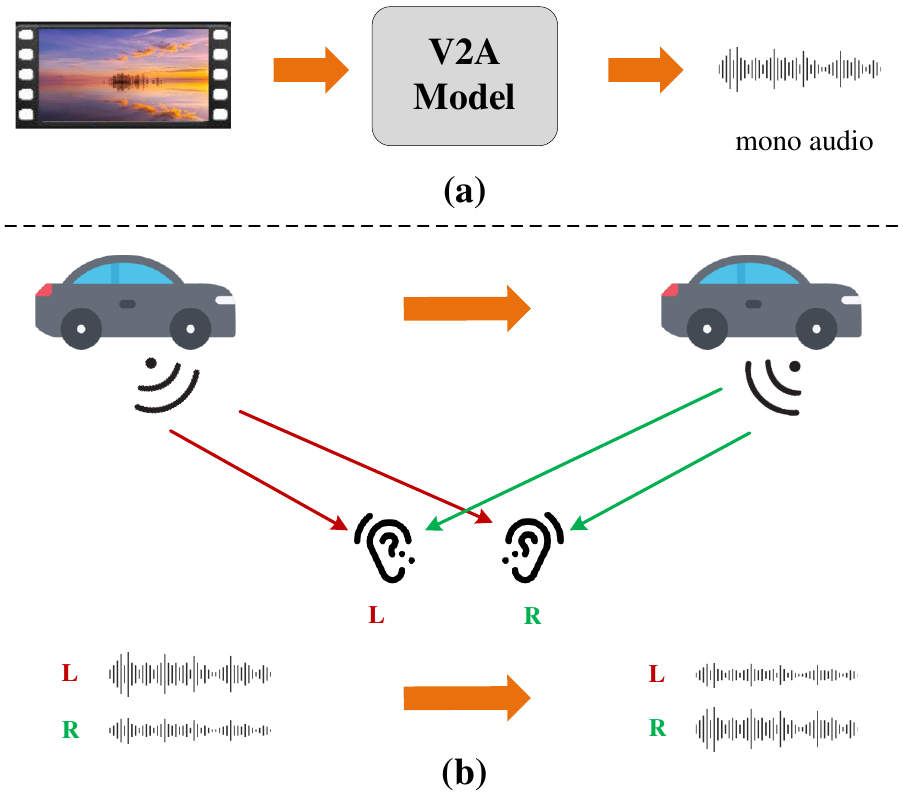}
    \caption{(a) The existing V2A model generates audio without spatial information. (b) There are differences in the audio information transmitted to both ears from sources at different positions, mainly reflected in the variations in volume and delay.}
    \label{fig:pre_figure}
\end{figure}
Driven by the advancement of generative artificial intelligence, recent studies have made further progress in modeling temporal consistency between visual and auditory modalities.
For example, DiffFoley \cite{luo2023diff} obtains visual representations aligned with the audio modality through contrastive audio-visual pretraining, and then utilizes these temporally aligned visual features to guide the diffusion model in generating monophonic audio.
Subsequently, Frieren \cite{wang2024frieren} applied rectified flow matching \cite{liu2022flow} to the V2A task and combined it with one-step distillation techniques, significantly improving the quality and efficiency of audio generation. 
V2A-Mapper \cite{wang2024v2a} aligns the model's output embeddings with audio representations by establishing a mapping between CLIP \cite{radford2021learning} and CLAP \cite{elizalde2023clap}. These aligned embeddings are then utilized to guide the generative model, AudioLDM \cite{liu2023audioldm}, in producing audio that corresponds to the video content. MMAudio \cite{cheng2025mmaudio} is trained on a larger-scale dataset and introduces a conditional synchronization module, thereby achieving significant improvements in audio quality and temporal consistency. Recently, AudioX \cite{tian2025audiox} introduces a multi-modal masked training strategy that enables audio generation from visual inputs. ThinkSound \cite{liu2025thinksound} innovatively employs a multimodal large language model-based Chain-of-Thought reasoning framework, opening new pathways to enhance the performance of V2A methods.


However, most of the aforementioned V2A methods are limited to generating monaural audio that lacks spatial awareness. Although \cite{tian2025audiox, liu2025thinksound} achieved binaural audio generation by simply using stereo Variational Autoencoders (VAEs), they did not consider the spatial consistency between video and audio.
As shown in Fig. \ref{fig:pre_figure} (b), the human auditory system achieves spatial localization by analyzing the differences in sound waves received by the left and right ears.
Among these, the interaural level difference (ILD), which forms an energy contrast, and the interaural time difference (ITD), which causes a temporal phase difference, together constitute the core cues for the brain to judge the direction of the sound source \cite{rayleigh1875our, leng2022binauralgrad}.
Based on this, \cite{dagli2024see} attempts to generate stereo audio by combining image segmentation with Room Impulse Responses (RIRs) modeling, improving the spatial consistency between video and audio.
However, the spatial audio generated by this method has relatively low audio quality, and its spatial consistency still needs improvement. Furthermore, it is only applicable to static sound sources. This makes it difficult to synthesize dynamic sound fields with time-varying characteristics.

To address the aforementioned issues, we propose FoleySpace, a binaural spatial audio  generation framework that aligns binaural audio with video content in both semantic and spatial dimensions. Our contributions are summarized as follows:

\begin{itemize}
    \item \textbf{We propose a novel framework for the generation of binaural spatial audio from silent video inputs.}
    The framework uses the estimated sound source trajectory and the mono audio output from the pretrained V2A model as conditions to guide a diffusion model in generating binaural spatial audio that is aligned with the visual input in semantic, temporal, and spatial dimensions.

    \item \textbf{We propose a novel sound source localization method.} To achieve precise frame-wise localization of the sound source in the video, we employ an open-vocabulary object detection model and a monocular depth estimation model to estimate its spatial coordinates and depth in the video frames. 
    
    \item \textbf{We design a mapping mechanism from the visual plane to the 3D sound field.}  This mechanism can map sound source positions represented by 2D pixel coordinates in video frames into a 3D sound field coordinate system, providing a foundation for generating binaural audio that is spatially aligned with the visual content. Additionally, we smooth the estimated source trajectory to enhance the naturalness and smoothness of the generated audio.

	\item \textbf{The proposed method demonstrates the capability to generate moving sound sources.} We constructed a binaural audio training dataset based on recorded Head-Related Impulse Responses (HRIRs). This dataset includes dynamic sound source trajectories with smooth variations, enabling the model to produce binaural audio with naturally varying sound fields. User study shows that our method achieves better spatial consistency than existing approaches.

\end{itemize}

The remainder of the paper is organized as follows. Section \ref{sec: rel} reviews related work. Section \ref{sec:method} details the proposed FoleySpace and its two implementations. Section \ref{sec:dataset} describes the construction of the training dataset. Section \ref{sec:experi} reports the experimental results, including both subjective and objective evaluations. Finally, Section \ref{sec:con} concludes the paper.

\section{Related Work}
\label{sec: rel}

\subsection{Audio Spatialization}
The audio spatialization task aims to leverage visual information to recover spatially immersive binaural audio from channel-mixed audio.
For example, \cite{gao20192} decodes mixed audio into its corresponding binaural counterpart by integrating visual information about object and scene layouts into a convolutional neural network.
Following this, a multi-task framework \cite{garg2023visually} is developed to learn geometry-aware features for audio spatialization by considering the underlying room impulse responses, the consistency between visual streams and sound source locations, as well as the temporal geometric consistency of sounding objects.
Liu \textit{et al.} employed two audio decoders to generate the left and right channel audio outputs separately \cite{liu2024visually}. They also proposed an audio-visual matching loss to further explore the correlation between binaural audio and scene visual inputs.
SAGM \cite{li2024cross} trains the generator and discriminator by leveraging shared spatiotemporal visual information. During adversarial training, the two modules are dynamically optimized via alternating parameter updates, facilitating knowledge exchange and feature sharing.

However, these methods are limited to recovering binaural audio from mixed signals and cannot be generalized to arbitrary monaural audio.



\begin{figure*}[t]
    \centering
    \includegraphics[width=0.88\textwidth]{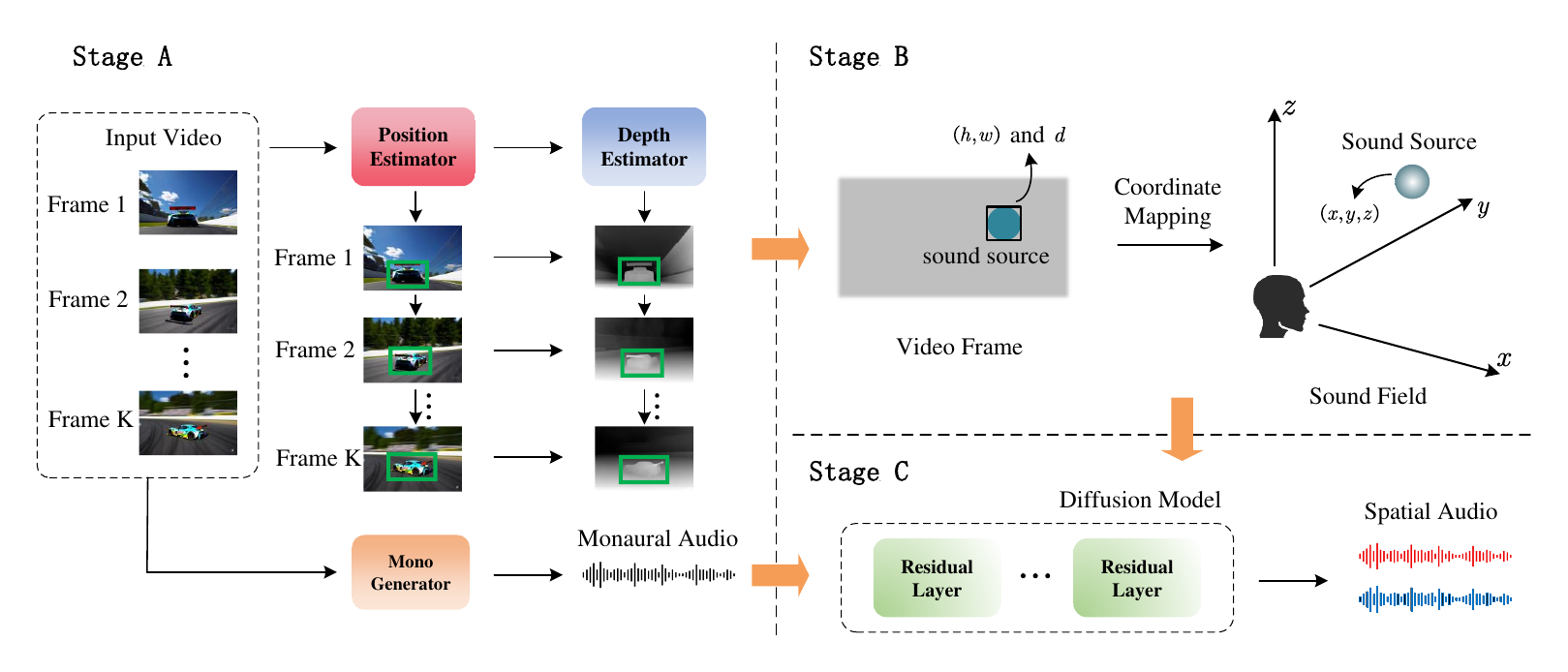}
    \caption{{Overview of the proposed FoleySpace framework, consisting of three stages. \textbf{Stage A}: The position estimator estimates the location of the sound source in each frame, while the depth estimator estimates the depth of the sound source. Meanwhile, the video signal is input to the mono generator to produce monaural audio. 
    \textbf{Stage B}: The information estimated by the position and depth estimators is then mapped to the 3D sound field, forming the motion trajectory of the sound source.
    \textbf{Stage C}: After receiving monaural audio and the 3D motion trajectory of the sound source as conditions, the diffusion model generates binaural spatial audio that matches the sound source direction in the video.}}
    \label{fig:overview}
\end{figure*}

\subsection{First-order Ambisonics Generation}
The First-order Ambisonics (FOA) Generation generation method based on diffusion models has been investigated in [11–13]. ImmerseDiffusion~\cite{heydari2024immersediffusion} takes text descriptions and spatial parameters as inputs to generate the aligned FOA audio. Diff-SAGe \cite{kushwaha2024diff} introduces a multi-condition encoder that integrates input conditions into a unified representation, thereby guiding the noise generation to produce FOA waveforms that match the given text description and spatial parameters. VISAGE \cite{kimvisage} combines camera parameters with visual features extracted using the CLIP \cite{radford2021learning} encoder to generate FOA from Field-of-View (FoV) format videos. Building upon this, \cite{Liu2025OmniAudio} proposes a dual-branch model architecture specifically designed for synthesizing FOA audio in 360$^\circ$ videos.

Unlike the above methods, we focus on generating binaural audio from common 2D videos. The proposed method does not rely on external parameters such as spatial coordinates or camera orientation, but instead guides binaural audio generation using positional information estimated from the video.

\section{FoleySpace}
\label{sec:method}
This section describes the proposed FoleySpace framework, which comprises sound source position estimator, depth estimator, coordinate mapping, and the diffusion model used to generate binaural audio. The overview of this method is shown in Fig. \ref{fig:overview}.

\subsection{Sound Source Position Estimator}
To obtain the trajectory of the sound source in 3D space, we first need to locate its coordinates in the 2D video. 
Traditional closed-set object detection models are limited by predefined fixed category labels and struggle to address the need for detecting objects of unknown categories. 
To address this issue, we utilize an open-vocabulary object detection model, YOLO-World \cite{cheng2024yolo}, as the position estimator in our framework. This model can achieve zero-shot localization of objects in 2D video. The position estimator takes text labels and video frames as input to obtain the source’s ground-truth bounding box. In addition, an offline vocabulary can be set up to allow inference without text input. The model assigns a confidence score to each label in the offline vocabulary based on the video content, and then we select the label with the highest confidence score as the actual input label.


We locate the sound source objects frame by frame in the video to obtain the corresponding bounding boxes, and calculate their geometric centers to determine their positions in the 2D image. For the $k$-th frame of the video, the planar coordinates  of the sound source are defined as follows:
\begin{equation}
\left( w_k, h_k \right) ,\quad {\forall} k= 1,2,\cdots ,K,
\label{planar_coordinate}
\end{equation}
where $w_k$ and $h_k$ denote the horizontal and vertical coordinates, respectively, of the center pixel of the bounding box in the {$k$}-th frame. $K$ is the total number of frames in the video.

\subsection{Sound Source Depth Estimator}
We use the monocular depth estimation model DepthMaster \cite{song2025depthmaster} to estimate the depth information of the sound source in the image.
For the input image $\mathbf{I}\in \mathbb{R} ^{\mathrm{H}\times \mathrm{W}\times 3}$, the output of DepthMaster is a depth map $\mathbf{D}\in \mathbb{R} ^{{H}\times {W}}$, where $H$ and $W$ represent the height and width of the image, respectively.
Each element $\mathbf{D}_{i,j}$ represents the estimated depth value of the pixel at position $\left( {i},{j} \right) $ in the image. Therefore, the depth value of the sound source coordinate corresponding to the $k$-th frame in Eq. \eqref{planar_coordinate} is defined as follows:
\begin{equation}
	d_k = \mathbf{D}_{h_k, w_k}, \quad \forall k = 1, 2, \dots, K.
\end{equation}

\subsection{Mapping from 2D Visual Plane to 3D Sound Field}
\label{mapping}
After obtaining the coordinates and depth of the sound source in the video frame, we map them to 3D spatial coordinates that align with the visual content, as shown in Stage B of Fig. \ref{fig:overview}.

To ensure consistent mapping for images with different resolutions, the mapping factor is determined by the width of the video frame, which is defined as:
\begin{equation}
\delta = \frac{2S_y}{W},
\label{eq:delta}
\end{equation}
{where
$S_y$ denotes} the maximum absolute distance of the sound source along the 
$y$-axis relative to the origin (audience) in the 3D sound field.

Then, at the $k$-th frame, the mapping from planar coordinates $\left( {h}_{{k}},{w}_{{k}},{d}_{{k}} \right)$ to the 3D sound field coordinates  centered on the audience $\left( x_k,y_k,z_k \right)$ can be expressed as
\begin{equation}
\left\{\begin{aligned}
x_k&=\delta \tilde{d}_k,\\
y_k&=\delta (w_k-\small{\frac{W}{2}}),\\
z_k&=-\delta (h_k-\frac{H}{2}),
\end{aligned}\right.
\end{equation}
where $\tilde{d}_k$ is a pixel depth value that has been normalized and scaled, defined as:
\begin{equation}
\tilde{d}_k = \gamma\left( \frac{d_k - \min(\mathbf{D})}{\max(\mathbf{D}) - \min(\mathbf{D})} \right).
\end{equation}
Here, we apply min-max normalization to convert the relative depth $d_k$ into absolute depth, as in \cite{dagli2024see}.
We also introduce a scaling factor $\gamma = \frac{W}{2}$  to adjust the distance between the sound source and the listener, thereby simulating a reasonable listening distance during video watching.
Finally, the motion trajectory of the sound source in 3D space can be formulated as follows
\begin{equation}
\mathcal{T} = \{ (x_k, y_k, z_k) \}_{k=1}^{K}.
\end{equation}
This mapping mechanism aligns the image center with the listener’s position in the 3D sound field, thereby ensuring spatial consistency between the visual content and the audio.

Finally, to improve the naturalness and fluency of the generated audio, we need to smooth the estimated sound source trajectory.
We first compute the frame-wise 3D motion magnitude:
\begin{equation}
\Delta_k = \left\| (x_{k+1}, y_{k+1}, z_{k+1}) - (x_k, y_k, z_k) \right\|_2.
\end{equation}
We consider values of $\Delta_k$ exceeding the 95$\%$ threshold as outliers and discard the estimated coordinates of the corresponding frames and their adjacent frames. Subsequently, we perform linear interpolation on each missing coordinate dimension separately to reconstruct the complete trajectory. The resulting smoothed 3D motion trajectory of the sound source is denoted as $\tilde{\mathcal{T}}$.

\begin{figure}[t]
    \centering
    \includegraphics[width=0.90\columnwidth]{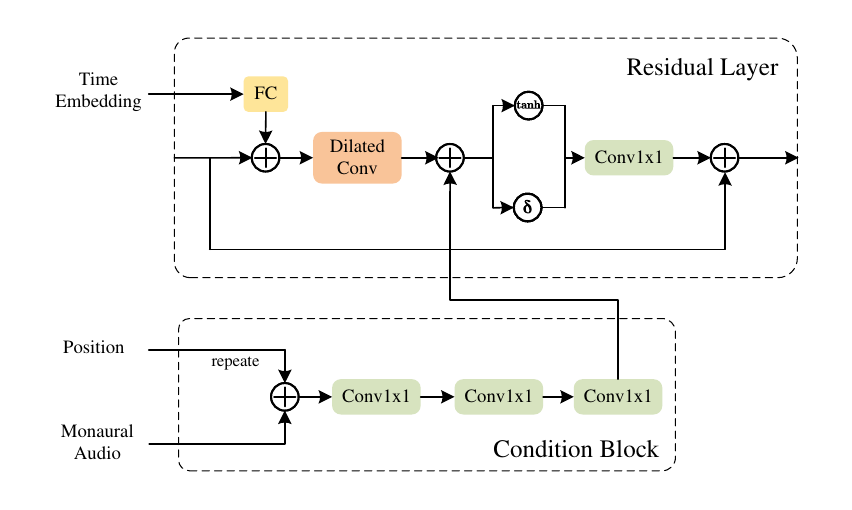}
    \caption{{Illustration of the architecture of the residual layer and the proposed condition block in the diffusion model. }}
    \label{fig:diffusion}
\end{figure}

\subsection{Diffusion-Based Binaural Spatial Audio Generation}
After obtaining the sound source trajectory, we use it as a condition, together with the monaural audio generated by the pretrained mono generator, to guide the diffusion model in generating binaural audio that is spatially aligned with the video.

We utilize MMAudio~\cite{cheng2025mmaudio} as the mono generator in the proposed framework, which can generate mono audio $s_{n}^{mono}$ that is semantically aligned but lacks spatial information. Therefore, the condition $c$ of the diffusion model consists of two parts: $s_{n}^{mono}$ and $\tilde{\mathcal{T}}$.
We design a condition block to integrate the information from the two components and subsequently use its output as a conditioning signal to guide the generation process of the diffusion model, as shown in Fig. \ref{fig:diffusion}. We replicate $\tilde{\mathcal{T}}$ along the time dimension to match the length of $s_{n}^{mono}$, then concatenate both along the channel dimension, and finally input the resulting tensor into the backbone network through a series of convolution operations.

We use DiffWave \cite{kong2020diffwave} as the backbone for the diffusion module. It is capable of generating high-quality audio waveforms, and significantly improves generation efficiency due to its bidirectional dilated convolution and fast sampling algorithm.
To support binaural audio processing, we adjusted the number of channels in the model’s input and output convolutional layers and correspondingly modified the condition block as mentioned above.

\begin{figure}[t]
    \centering
    \vspace{0.5cm}
    \includegraphics[width=0.98\columnwidth]{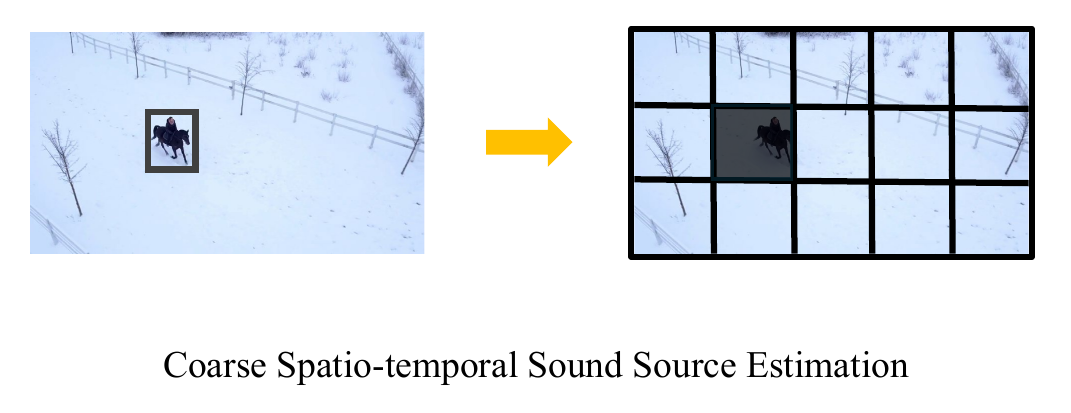}
    \caption{Illustration of coarse-grained spatio-temporal sound source estimation. It first detects the bounding box of the sound source, and then determines the grid cell in which it is located.}
    \label{fig: coarse}
\end{figure}

\subsection{Coarse-grained Spatio-temporal Implementation of FoleySpace}
In this subsection, we propose a coarse-grained spatio-temporal implementation of FoleySpace. 

For sound source estimation, we divide the video plane into a 5×3 grid, resulting in 15 cells. The grid cell containing the sound source is estimated once every second. Specifically, as shown in Fig. \ref{fig: coarse}, the scheme first determines which cell contains the center of the sound source's bounding box:
$\hat{s}_k \in \mathcal{C}_{i, j},$
and then take the geometric center of the cell \(\mathcal{C}_{i, j}\) as the final estimate of the source location, where $i = 1, \ldots, 5$ indexes the five horizontal grid columns, and $j = 1, \ldots, 3$ indexes the three vertical grid rows. Therefore,  for this scheme, $(h, w)$ in Eq. {\eqref{planar_coordinate}} corresponds to the coordinates of the center pixel of the grid cell containing the sound source, and $K$ represents the video duration in seconds, with estimations performed at a rate of one frame per second. 

For depth estimation, we also discretize the estimated source depth into five evenly spaced values and map them, together with the spatial coordinates, into the 3D sound field. This results in a total of $5 \times 3 \times 5$ spatial positions, which can be regarded as a sampled version of the fine-grained spatio-temporal scheme. Therefore, these directional cues are still input into the conditional branch of the diffusion model in the same way, guiding the model to generate binaural spatial audio corresponding to the source trajectory.

Compared with the fine-grained spatio-temporal scheme, the sound source movement trajectory generated by this scheme is more discrete, facilitating a quantitative evaluation of the accuracy of the source estimation method.

\begin{figure}[t]
    \centering
    \includegraphics[width=\columnwidth]{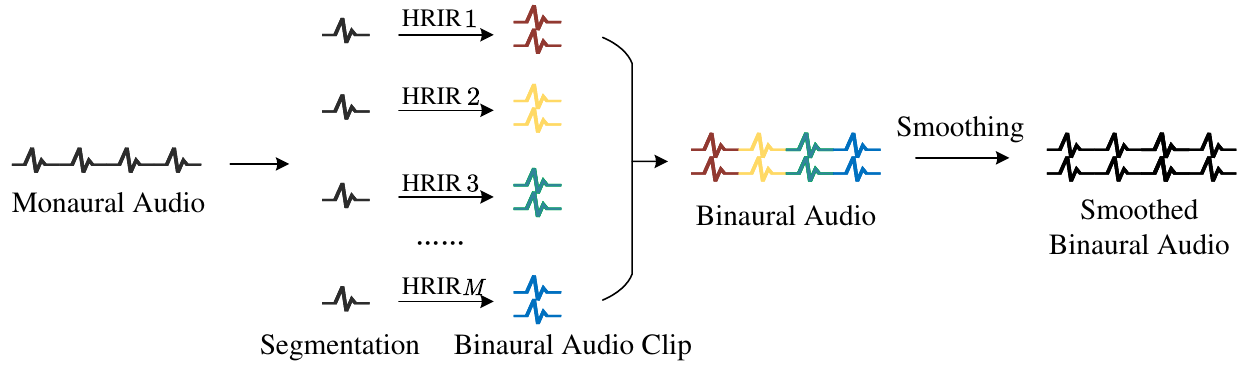}
    \caption{{Construction of binaural audio for moving sound sources. Each HRIR represents the direction of the sound source at a specific moment along its trajectory.}}
    \label{fig:dataset}
\end{figure}

\section{{Training Set Construction of Binaural Audio Data}}
\label{sec:dataset}
We randomly sample 10,000 monaural audio clips from the VGGSound dataset \cite{chen2020vggsound}, each segmented into 8-second segments, to serve as the foundational data for constructing the training set of the diffusion model.
We utilize the recorded HRIRs from ten subjects (pp87 to pp96) in the HUTUBS head-related transfer function (HRTF) database \cite{brinkmann2019hutubs} to process these audio clips, thereby generating corresponding binaural audio.

The database provides measurements at 440 HRIRs positions distributed over a spherical grid. Based on the mapping relationship, in the fine-grained spatio-temporal scheme, we select HRIRs with azimuth angles in the range of [90°, 270°] and elevation angles in the range of [-40°, 40°], both of which include the 0° reference point. In the coarse-grained spatio-temporal scheme, we select HRIRs for 15 fixed directions, corresponding to combinations of 5 azimuth angles ($80^\circ$, $40^\circ$, $0^\circ$, $320^\circ$, and $280^\circ$) and 3 elevation angles ($40^\circ$, $0^\circ$, and $-40^\circ$). Then, we construct the training dataset based on the following steps.


\subsection{Distance-Varying HRIRs}
\label{dataset: dis}
Since the recorded HRIRs are at a fixed distance relative to the listener (1.47m), we introduce varying propagation delays through time-domain resampling \cite{martin2007interpolation} to simulate the effect of different source distances. This process stretches or compresses the temporal structure of HRIRs in a linear fashion, thereby approximating the shift in time-of-arrival introduced by sound traveling over varying distances. 
By resampling, the HRIRs can accurately model the time-of-arrival characteristics of sound sources within 0–6m. For the coarse-grained spatio-temporal scheme, we fix the depth scaling at five distances {of $\{1,2,3,4,5\}$ meters respectively}.

\subsection{Moving Sound Source Simulation}
\label{dataset: mov}

To simulate the movement of the sound source in 3D space, we uniformly divided the monaural audio provided by VGGSound into $M$ segments. 
These segments were then convolved with the processed HRIRs (as described in Section \ref{dataset: dis}) corresponding to each position along the sound source’s motion trajectory, generating binaural spatial audio segments for each location. $M$ denotes the direction variation rate of the sound source position. Fig. \ref{fig:dataset} illustrates the process of constructing binaural audio with moving sound sources.

To ensure perceptual continuity and realistic spatial transitions across the simulated path, we further applied the spatial smoothing approach used in \cite{diaz2021gpurir} across the $M$ binaural audio segments. 
Specifically, for each binaural audio segment, an index-ordered weight was assigned based on the normalized motion trajectory of the sound source. The smoothed binaural audio segement $\tilde{\mathbf{x}}_n$ is given by:
\begin{equation}
\tilde{\mathbf{x}}^{(m)}_n = (1 - \alpha_n) {\mathbf{x}}^{(m)}_n + \alpha_n{\mathbf{x}}^{(m+1)}_n, \quad {\forall}i = 1, 2, \ldots, M,
\end{equation}
where \( \alpha_n\in[0, 1] \) is a linearly increasing interpolation weight that transitions from 0 to 1 across the segment between positions \( m \) and \( m{+}1 \); \(\mathbf{x}^{(m)}_n\) and \(\mathbf{x}^{(m+1)}_n\) denote the binaural audio segement corresponding to spatial positions \(m\) and \(m{+}1\), respectively. 

For the fine-grained spatio-temporal scheme, we set the direction variation rate $M$ to 200, corresponding to a minimum interval of 0.04 seconds for sound source position changes. 
As for the coarse-grained spatio-temporal scheme, we set $M = 8$, corresponding to a sound source direction change occurring once per second. The complete binaural audio is finally constructed by concatenating all the smoothed segments in temporal order.

\subsection{Binaural Audio Dataset Construction}
We evenly divided the 10,000 monaural audio clips from the VGGSound dataset into 10 groups, with each group corresponding to one subject's HRIRs.

For each monaural audio input, we randomly generate a source movement trajectory and synthesize corresponding binaural spatial audio using the method in Section \ref{dataset: mov}. This approach yields 10,000 spatial audio samples per scheme (fine/coarse-grained spatio-temporal scheme) for training their respective diffusion models.

\section{Experiments}
\label{sec:experi}
\subsection{Dataset}
The training dataset for the diffusion models  construted in Section~\ref{sec:dataset}. 
We carefully selected 9,212 single-source audio samples from the VGGSound \cite{chen2020vggsound} dataset and manually annotated their source locations in a grid format (as shown in Fig. \ref{fig: coarse}) to construct the evaluation dataset VGGSound-Solo~\cite{VGGSoundSolo2025}.
The training and evaluation sets do not overlap.

\subsection{Baselines}
We select the following methods as our baselines:
\begin{itemize}
\item \textbf{MMAudio} \cite{cheng2025mmaudio}:   A V2A method that generates monaural audio lacking spatial characteristics. This method also serves as the mono generator component in our framework.

\item \textbf{See2Sound} \cite{dagli2024see}: It uses a semantic segmentation model to identify visually interesting regions, and then synthesizes 5.1 surround spatial audio using the Image Source Method \cite{allen1979image}.

\item \textbf{AudioX} \cite{tian2025audiox}: It generates audio from video through multimodal masked training and leverages a pre-trained stereo VAE to endow the generated audio with a sense of spatiality.
\item \textbf{ThinkSound} \cite{liu2025thinksound}: It is a contemporaneous work that employs a chain-of-thought reasoning framework based on multimodal large language models and also incorporates a stereo VAE to generate binaural audio from video.

\end{itemize}

\subsection{Evaluation Metrics}

\subsubsection{User Study}
Considering the limitations of existing metrics in comprehensively evaluating the generated binaural spatial audio, we conducted a user study for subjective assessment.  
We collected a total of 32 video samples, covering both real and generated content, with sound sources that are either stationary or moving. For each video, we generated the corresponding audio using our method and the aforementioned baselines, and conducted a subjective evaluation with 24 participants. Each participant will evaluate a total of 192 video samples (32 videos {$\times$} 6 methods) and rate the audio component of each video from the following four perspectives:
\begin{itemize}
	\item \textbf{Perceived Spatiality Score (PSS)}: Evaluates the extent to which listeners perceive spatial characteristics in an audio sample. It reflects listeners’ sense of spatial separation, directionality, depth, and immersive quality of the sound field.

	\item \textbf{Spatial Alignment (SA)}: Assesses whether the spatial cues in the audio—such as the direction and distance of the sound source—are consistent with the visual information in the video, providing a realistic sense of space.

	\item \textbf{Temporal Alignment (TA)}: Measures the synchronization between audio and video over time, ensuring that sounds occur at the correct moments and align well with visual actions.

	\item \textbf{Semantic Consistency (SC)}: Evaluates how well the audio content matches the video semantically, for example, whether the type of sound is appropriate and accurately reflects the events or actions shown.
    
	\item \textbf{Audio Quality (AQ)}: Evaluates the clarity, naturalness, and any noise or distortion in the audio alone, without considering the video content.

\end{itemize}
Participants are asked to rate each video on a {one-to-five} scale for each evaluation dimension, with higher scores indicating better performance. Detailed scoring criteria for each dimension are provided in the Appendix. Subsequently, we computed the Mean Opinion Score (MOS) for each method. An example of the user study questionnaire is shown in Fig. \ref{fig:user_study} in the Appendix.

\subsubsection{Quantitative Evaluation}
To further improve the experiment, we incorporated objective metrics for evaluation.
Following \cite{cheng2025mmaudio}, we evaluate the similarity between the feature distributions of generated and real audio using the Fréchet Distance (FD) and Kullback–Leibler (KL) divergence. For FD, we adopt PANNs \cite{kong2020panns} and VGGish \cite{gemmeke2017audio} as feature extractors, with the resulting metrics  $\text{FD}_\text{PANNs}$ and $\text{FD}_\text{VGG}$, respectively. For KL divergence, we use PANNs and PaSST \cite{koutini2022efficient} as classifiers, resulting in metrics $\text{KL}_\text{PANNs}$ and $\text{KL}_\text{PaSST}$. We also adopt the Inception Score (IS)~\cite{salimans2016improved} to evaluate the quality of the generated audio. Furthermore, to evaluate potential semantic and temporal discrepancies between the generated binaural spatial audio and the video content, we adopt IB-score \cite{viertola2025temporally} and DeSync \cite{cheng2025mmaudio} as the evaluation metrics for semantic and temporal consistency, respectively.

Additionally, we use the Mean Absolute Error (MAE) metric, as in \cite{zhao2025dualspec}, to quantify the angular error between the sound source predicted by the position estimator and the ground truth. This metric is suitable for the coarse-grained spatio-temporal scheme. 
Specifically, through sound source estimation and mapping, we obtain the 3D trajectory of the sound source as described in Section \ref{mapping} and compute its azimuth and elevation angles at each moment. Similarly, the ground truth positions\footnote{The ground truth positions of the sound sources in video frames are manually annotated and provided in the VGGSound-Solo.} of the sound source in video frames are mapped to 3D space, and their corresponding azimuth and elevation angles are calculated.
Then, the MAE between the predicted and ground truth values is calculated separately, denoted as $\text{MAE}_\alpha$ for azimuth and $\text{MAE}_\varepsilon$ for elevation. Considering the periodicity, $\text{MAE}_\alpha$ is calculated as follows:
\begin{equation}
\mathrm{MAE}_{\alpha}=\sum_{n=1}^N{\min \left( \left| \hat{\theta}^{\alpha}_n-\theta ^{\alpha}_n \right|,360-\left| \hat{\theta}^{\alpha}_n-\theta^{\alpha}_n \right| \right)},
\end{equation}
where $\hat{\theta}^{\alpha}_n$ and $\theta ^{\alpha}_n$ represent the estimated azimuth angle and the actual azimuth angle, respectively. And $\text{MAE}_\varepsilon$ is formulated as:
\begin{equation}
\mathrm{MAE}_{\varepsilon}=\sum_{n=1}^N{\min \left( \left| \hat{\theta}^{\varepsilon}_n -\theta^{\varepsilon}_n \right| \right)},
\end{equation}
where $\hat{\theta}^{\varepsilon}_n$ and $\theta ^{\varepsilon}_n$ represent the estimated elevation angle and the actual elevation angle, respectively.
\begin{table*}[h]
    \centering
    \caption{User study results of different methods. * denotes the mono generator in the proposed framework. }
    \setlength{\tabcolsep}{3.9mm}
    \begin{tabular}{cccccccc}
        \toprule
        Audio Type & Method &  PSS $\uparrow$ & SA $\uparrow$    & TA $\uparrow$ & SC $\uparrow$ & AQ $\uparrow$  \\
        \midrule
            Monaural & MMAudio{$^*$}~\cite{cheng2025mmaudio} & \textcolor{gray}{$1.54\pm1.11$} & \textcolor{gray}{$1.67\pm1.03$} &\textcolor{gray}{${3.86\pm0.93}$} & \textcolor{gray}{${3.77\pm0.99}$}& \textcolor{gray}{$\boldsymbol{3.72\pm0.98}$}\\\hline
           \multirow{5}{*}{Multichannel} & 
            See2Sound~\cite{dagli2024see} & $1.32\pm0.64$ & $1.25\pm0.58$& $1.45\pm1.86$& $1.13\pm0.44$& $1.39\pm0.72$\\           
           & AudioX~\cite{tian2025audiox} & $2.76\pm1.13$ & $2.79\pm1.16$ &${2.77\pm1.21}$ & ${2.92\pm1.22}$& $2.90 \pm1.15$\\

            & ThinkSound~\cite{liu2025thinksound} & $2.47\pm1.20$ & $2.56\pm1.27$& $2.74\pm1.30$& $2.57\pm1.39$& $2.71\pm1.27$\\
            & FoleySpace (Ours) & $\boldsymbol{3.72\pm0.96}$ & $\boldsymbol{3.85\pm0.97}$& $\boldsymbol{3.79\pm0.96}$& ${3.81\pm0.99}$&
            ${3.69\pm1.08}$\\
            
            & FoleySpace\_coarse (Ours) & ${3.67\pm1.02}$ & ${3.79\pm1.03}$& $3.71\pm0.93$& $\boldsymbol{3.92\pm1.00}$&$ \boldsymbol{3.71\pm0.96}$\\
        \bottomrule
    \end{tabular}
    \label{tab:compar_subj}
\end{table*}

\begin{table*}[h]
\centering
\caption{Objective evaluation results for VGGSound-Solo. * denotes the mono generator in the proposed framework.}
\setlength{\tabcolsep}{3.27mm}
\begin{tabular}{ccccccccc}
\toprule
Audio Type & Method   & $\text{FD}_{\text{PANNs}} \downarrow$ & $\text{FD}_\text{VGG} \downarrow$  & $\text{KL}_\text{PANNs} \downarrow$ & $\text{KL}_\text{PaSST} \downarrow$ & IS $\uparrow$  & IB-score $\uparrow$ & DeSync $\downarrow$\\
\midrule
Monaural & MMAudio$^{*}$ \cite{cheng2025mmaudio}  & \textcolor{gray}{4.57} & \textcolor{gray}{0.94}&  \textcolor{gray}{1.73}& \textcolor{gray}{1.48}& \textcolor{gray}{18.32}& \textcolor{gray}{31.74}& \textcolor{gray}{0.46}\\\hline

\multirow{5}{*}{Multichannel} & See2Sound \cite{dagli2024see} &  51.19 &  9.21   & 4.34  &   3.91& 3.42& 7.95&  1.28 \\
& AudioX \cite{tian2025audiox}  &   13.54 &  1.61    &  2.78 &  2.67&  14.35& 24.08 & 1.23 \\

& ThinkSound \cite{liu2025thinksound} & 8.42 & \textbf{1.40} & 1.95 &1.75 & 11.57  &23.32 & 0.57  \\
& FoleySpace (Ours)  & {7.25} & 1.73   & \textbf{1.76} & {1.45} & {14.55} &  {26.42} & \textbf{0.50} \\
& FoleySpace\_coarse (Ours)  & \textbf{6.81} & 1.58   & {1.77} & \textbf{1.43} & \textbf{15.24} &  \textbf{26.71} & {0.51} \\
\bottomrule
\end{tabular}
\label{tab:your_label}
\end{table*}

\begin{table}[ht]
\centering
\setlength{\tabcolsep}{1.6mm}
\caption{The effect of text labels on sound source estimation accuracy for VGGSound-Solo inference. }
\begin{tabular}{cccccc}
\toprule
\multirow{2}{*}{Metric} & \multirow{2}{*}{Text Label} & \multicolumn{4}{c}{Offline Vocabulary} \\
\cline{3-6}
& & COCO & ImageNet-1K& VGGSound & AudioSet  \\
\midrule
$\mathrm{MAE}_{\alpha}$ & 39.62& 48.75& 48.82& 49.61& 45.04\\
$\mathrm{MAE}_{\varepsilon}$ & 13.01& 16.01& 18.64& 16.21& 15.86\\
\bottomrule
\label{tab:text}
\end{tabular}
\end{table}

\begin{table}[h]
\centering
\setlength{\tabcolsep}{1.35mm}
\caption{Comparison of binaural audio quality based on RIRs and HRIRs for VGGSound-Solo.}
\label{tab:comp_phsic}
\begin{tabular}{ccccccc}
\toprule
Scheme&{Type}& $\text{FD}_{\text{PANNs}} \downarrow$ & $\text{FD}_\text{VGG} \downarrow$ & $\text{KL}_\text{PANNs} \downarrow$ & $\text{KL}_\text{PaSST} \downarrow$ & IS $\uparrow$ \\
\midrule
\multirow{2}{*}{Fine} &RIRs  & 69.03 &  12.37 & 3.78& 3.22& 2.05\\
&HRIRs & \textbf{7.25} & \textbf{1.73} & \textbf{1.76} & \textbf{1.45}  & \textbf{14.55}\\ \hline
\multirow{2}{*}{Coarse} &RIRs  & 62.81 &  11.14& 3.68& 3.14& 2.31\\
&HRIRs & \textbf{6.81} & \textbf{1.58} & \textbf{1.77} & \textbf{1.43}  & \textbf{15.24}\\
\bottomrule
\label{tab:rir}
\end{tabular}
\end{table}

\subsection{Implementation}
In our framework, the video frame rate was set to 25 fps. The distance $S_y$ was set to 1.47 m. We adopted DiffWave \cite{kong2020diffwave} as the backbone of our diffusion module for generating binaural spatial audio. The number of residual layers, the number of residual channels, and the dilation cycle length were set to 60, 128, and 10, respectively. We trained the diffusion model for 200K steps using a batch size of 12 on 4 NVIDIA H100 GPUs. During training, the noise followed a linear schedule from $1\mathrm{e}{-4}$ to $0.02$ over 200 steps. For inference, we set the inference step to 50 to balance efficiency and fidelity.

Two diffusion models were trained, one targeting the fine-grained spatio-temporal scheme and the other the coarse-grained scheme, and were referred to as FoleySpace and FoleySpace\_coarse, respectively.

\subsection{Subjective Evaluation Results}
{Table} \ref{tab:compar_subj} compares the results of FoleySpace with those of the four baseline methods in the user study.
Compared to MMAudio which generates monaural audio, our approach demonstrates significant improvements in both PSS and SA metrics, indicating that the binaural spatial audio generated by FoleySpace provides clearer spatial orientation compared to monophonic audio. Furthermore, the proposed method shows performance comparable to that of MMAudio in TA, SC, and AQ metrics, suggesting that FoleySpace effectively preserves the original audio quality while achieving enhanced spatial perception.

Compared to the other three multi-channel audio generation methods, the proposed approach demonstrates superior performance across all evaluation metrics. This indicates that FoleySpace holds significant advantages in terms of audio quality, spatial perception, and the semantic, temporal, and spatial alignment between audio and visual modalities. Although both AudioX and ThinkSound employ stereo VAEs to generate binaural audio, they do not specifically focus on optimizing spatial audio generation or achieving consistent spatial alignment between audio-visual modalities. Thus, their performance in these aspects lags behind.
Interestingly, See2Sound, as a spatial audio generation method, underperformed compared to the monophonic audio generation method MMAudio on both the PSS and SA metrics. The reason may lie in the fact that See2Sound attempts to identify multiple regions of interest, generate multiple audio signals, and simply superimpose them. This process leads to auditory clutter, which consequently degrades its performance on metrics like PSS and SA.

\begin{figure*}[t]
    \centering
    \includegraphics[width=0.86\textwidth]{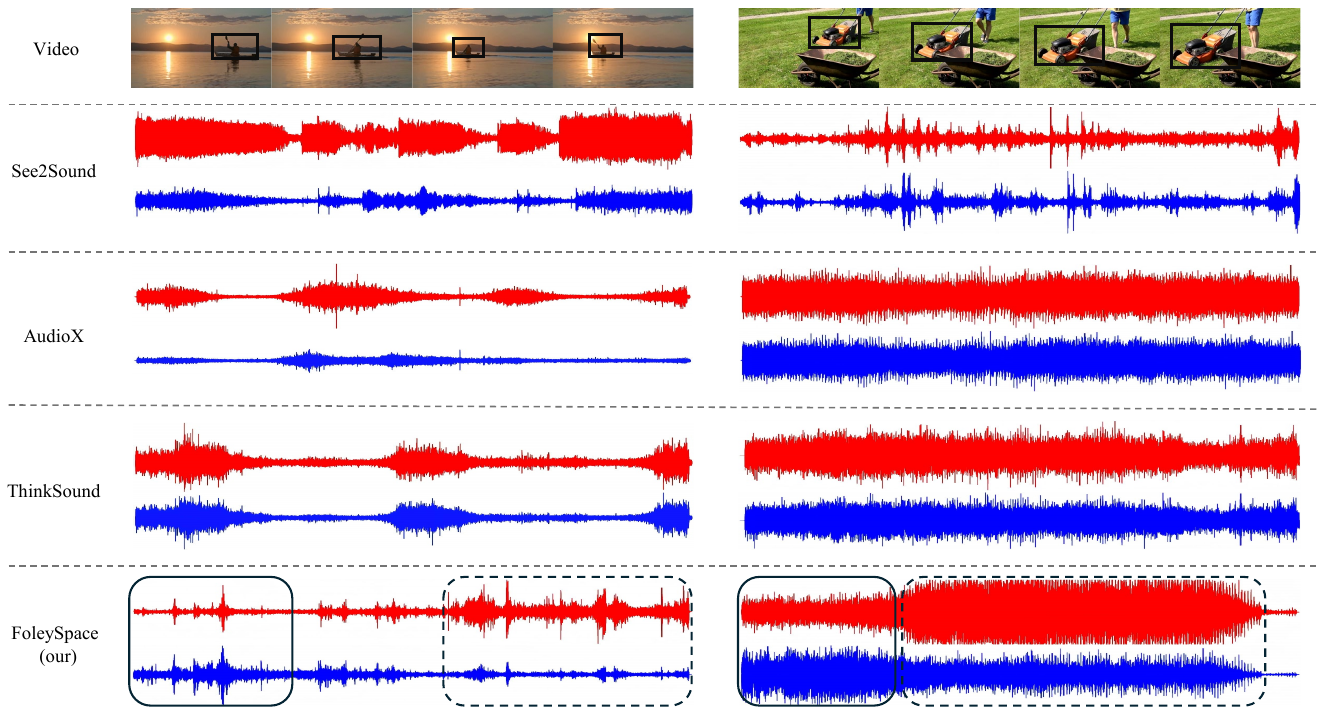}
    \caption{Ilustration of binaural audio waveforms compared with baselines. Red represents the left-ear audio, and blue represents the right-channel audio. For FoleySpace, solid boxes indicate that the right-channel audio is louder than the left-channel audio, which means the sound source is located on the right side of the visual scene; dashed boxes indicate that the right-channel audio is quieter than the left-channel audio, which means the sound source is located on the left side of the scene.}
    \label{fig:compare_wave}
\end{figure*}

\subsection{Quantitative Results}

\subsubsection{Numerical Results}
Table 2 presents a comparison of objective metrics across various methods on the VGGSound-Solo evaluation set.
The results in the table indicate that among multichannel audio generation methods, our approach ranks first overall, with our performance only slightly lower than that of ThinkSound in $\text{FD}_\text{VGG}$. Compared to the monaural audio generation method MMAudio, our approach slightly lags behind in objective metrics, but the gap is not significant. This result aligns with the conclusions of subjective evaluations, fully validating the effectiveness of the proposed spatial audio generation framework.

\subsubsection{Visualization of Audio-Visual Spatial Consistency}
Figure \ref{fig:compare_wave} visually compares how well the spatial positions of audio generated by different methods match the sound source location in the video. The red waveform corresponds to the left-channel audio, while the blue waveform corresponds to the right-channel audio. It is worth noting that the audio generated by See2Sound  \cite{dagli2024see} is in $5.1$-channel surround sound, and the waveform shown in the figure is taken from its left and right channels.

In the first example, we can observe a character rowing from the right to the left side of the screen, while the corresponding sound source trajectory moves from left to right. Our method generates a waveform that clearly reflects this motion: in the initial stage, the louder volume in the right ear channel indicates the sound source is positioned on the right side of the screen; subsequently, the volume gradually shifts until the left ear channel becomes dominant, accurately corresponding to the movement of the sound source toward the left side of the screen. However, in ThinkSound, the dual-channel audio has identical volume levels, failing to convey a clear sense of direction. For See2Sound and AudioX, although there is a fixed difference in volume between the left and right channels, the lack of dynamic variation prevents the spatial direction of the generated audio from fully matching the actual position of the sound source in the video. This example result effectively demonstrates that FoleySpace outperforms other methods in spatial consistency between video and generated audio. The second example on the right side of Fig. \ref{fig:compare_wave} demonstrates the same result.

\subsection{Impact of Label Usage on Sound Source Estimation Performance}
Table \ref{tab:text} shows the impact of text labels on sound source estimation accuracy during inference. In the Table, the offline vocabulary assigns a confidence score to each built-in label based on the video content and selects the label with the highest score as the actual label for the sound source, thus eliminating the need for manual text labels during the inference stage. We adopted the built-in vocabulary in  YOLO-World, which is based on COCO dataset \cite{lin2014microsoft} category labels, and separately constructed vocabularies based on the labels from three datasets—ImageNet-1K \cite{deng2009imagenet}, VGGSound \cite{chen2020vggsound}, and AudioSet \cite{gemmeke2017audio}—for comparative analysis. Among these, COCO and ImageNet-1K are widely used image datasets, while VGGSound and AudioSet are sound event datasets. 

As shown in the table, the four different offline vocabulary lists achieve similar MAE performance. Compared to the method using text labels, they exhibit a difference of less than 10 degrees in $\mathrm{MAE}_{\alpha}$, while showing no significant difference in $\mathrm{MAE}_{\varepsilon}$. Since this difference is typically imperceptible to the human ear, employing an offline vocabulary for sound source estimation remains a straightforward yet effective approach.

\subsection{Comparison with RIRs-based Method}
Table \ref{tab:rir}  presents a comparison of binaural audio quality based on RIRs and HRIRs.
RIRs and HRIRs represent two different methods for constructing training data for  diffusion models. The diffusion model in our framework is trained on binaural audio generated by HRIRs simulations, with the data construction methodology detailed in Section \ref{sec:dataset}. RIRs characterize the acoustic transfer properties between a sound source and a receiver as impulse response sequences, simulating direct sound, reflections, and reverberation in a specific space. Similar to See2Sound \cite{dagli2024see}, we utilize Pyroomacoustics \cite{scheibler2018pyroomacoustics} to simulate RIRs-based binaural audio. Subsequently, we trained the diffusion model using synthesized binaural audio, and the results are shown in Table \ref{tab:rir}.

From the table, it is evident that for both fine-grained and coarse-grained spatio-temporal schemes, the HRIRs-based approach outperforms the RIRs-based approach in perceived audio quality. This indicates that the quality of binaural audio is highly dependent on the fidelity of spatial auditory cues. By accurately modeling the physiological filtering characteristics of the human ear, HRIRs can achieve a high level of audio realism.

\section{Conclusions}
\label{sec:con}
In conclusion, FoleySpace demonstrates the potential of generating binaural spatial audio from video and introduces a novel approach for modeling spatial consistency between video and audio. The method employs object detection for sound source estimation and designs a mapping mechanism to associate the 2D visual plane with the sound field, thereby enabling the generation of fine-grained 3D sound source trajectories. The diffusion model is used to generate binaural spatial audio guided jointly by the monaural audio and the sound source trajectories. Experimental results show that this method achieves performance comparable to state-of-the-art video-to-monaural-audio generation models, while surpassing existing spatial audio generation methods in spatial perception metrics.

\appendix
\label{append}

\begin{figure}[t]
    \centering
    \includegraphics[width=0.42\textwidth]{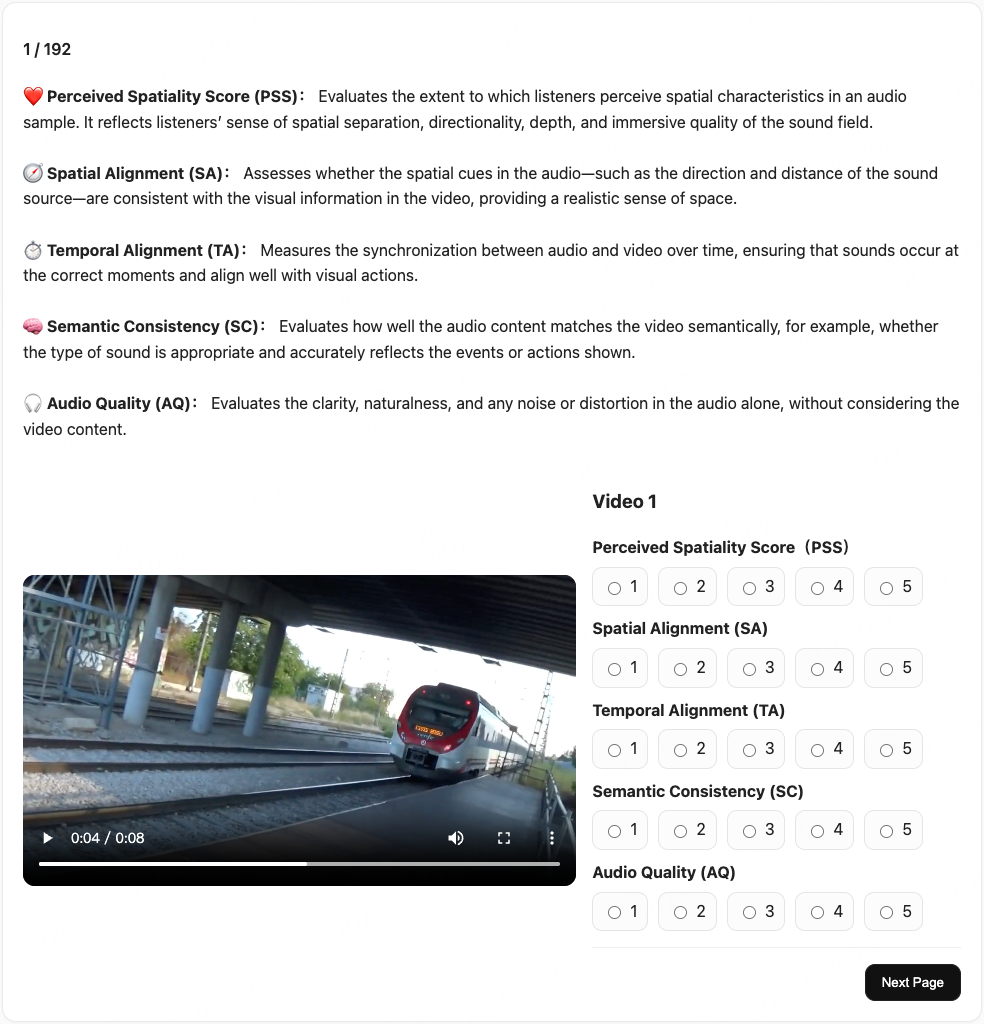}
    \caption{An example of the user study questionnaire. {The index of the current video is displayed in the upper-left corner, with 192 videos in total.}}
    \label{fig:user_study}
\end{figure}

This appendix provides an example of the user study questionnaire shown in Fig. \ref{fig:user_study} and a detailed explanation of the scoring criteria for each evaluation metric, as follows:

\begin{itemize}
\item \textbf{Perceived Spatiality Score (PSS)}: 
    \begin{itemize}[label=*]
    \item 1 - Almost no perceivable spatial characteristics; sound is concentrated in one     position or completely flattened, with no sense of direction, depth, or immersion.
    \item 2 - Weak sense of spatiality, unclear directionality or depth; sound source position is vague; very low sense of immersion.
    \item 3 - Some spatiality; able to perceive changes in direction and distance, but overall spatial separation or immersion is weak, and the soundstage is relatively flat.
    \item 4 - Clear spatial separation, directionality, and depth; strong sense of immersion, but with minor flaws in some sound source positions or details.
    \item 5 - Listeners clearly perceive spatial separation, directionality, depth, and stereoscopy; extremely strong sense of immersion; spatial reproduction feels natural and close to a real-life experience.
    \end{itemize}
\item \textbf{Spatial Alignment (SA)}: 
    \begin{itemize}[label=*]
    \item 1 - Completely inconsistent with the video’s spatial scene; direction is chaotic or lacks spatial sense.
    \item 2 - Weak sense of localization; direction and/or distance are seriously inconsistent.
    \item 3 - Basically reasonable, but somewhat vague or unstable.
    \item 4 - Accurate localization, largely matching the sound source position.
    \item 5 - Extremely high consistency, precise localization, and strong sense of immersion.
    \end{itemize}
\item \textbf{Temporal Alignment (TA)}: 
    \begin{itemize}[label=*]
    \item 1 - Severely advanced/delayed, completely out of sync.
    \item 2 - Delayed or unstable alignment, affecting the experience.
    \item 3 - Basically matched, with minor deviations.
    \item 4 - Good synchronization, slight deviation.
    \item 5 - Highly consistent, actions and sounds are synchronized.
    \end{itemize}
\item \textbf{Semantic Consistency (SC)}: 
    \begin{itemize}[label=*]
    \item 1 - Completely inconsistent with the video content, causing misleading semantics.
    \item 2 - Significant semantic deviation, unable to accurately convey the meaning.
    \item 3 - Roughly matched, with some ambiguity or minor deviations.
    \item 4 - Clearly expressed, basically consistent.
    \item 5 - Highly matched, accurate and natural.
    \end{itemize}
\item \textbf{Audio Quality (AQ)}: 
    \begin{itemize}[label=*]
    \item 1 - Severe distortion, obvious noise/breakage, seriously affecting comprehension.
    \item 2 - Noticeable noise, unnatural timbre, or heavy synthetic artifacts.
    \item 3 - Basically clear, with slight noise or mild synthetic artifacts, acceptable.
    \item 4 - Good, natural and smooth, with only minor flaws.
    \item 5 - Excellent, clear and natural, with no obvious flaws.
    \end{itemize}
    
\end{itemize}

\small
\bibliographystyle{IEEEtran}
\bibliography{Reference}

\begin{thebibliography}{10}
\providecommand{\url}[1]{#1}
\csname url@samestyle\endcsname
\providecommand{\newblock}{\relax}
\providecommand{\bibinfo}[2]{#2}
\providecommand{\BIBentrySTDinterwordspacing}{\spaceskip=0pt\relax}
\providecommand{\BIBentryALTinterwordstretchfactor}{4}
\providecommand{\BIBentryALTinterwordspacing}{\spaceskip=\fontdimen2\font plus
\BIBentryALTinterwordstretchfactor\fontdimen3\font minus \fontdimen4\font\relax}
\providecommand{\BIBforeignlanguage}[2]{{%
\expandafter\ifx\csname l@#1\endcsname\relax
\typeout{** WARNING: IEEEtran.bst: No hyphenation pattern has been}%
\typeout{** loaded for the language `#1'. Using the pattern for}%
\typeout{** the default language instead.}%
\else
\language=\csname l@#1\endcsname
\fi
#2}}
\providecommand{\BIBdecl}{\relax}
\BIBdecl

\bibitem{zhou2018visual}
Y.~Zhou, Z.~Wang, C.~Fang, T.~Bui, and T.~L. Berg, ``Visual to sound: Generating natural sound for videos in the wild,'' in \emph{Proceedings of the IEEE conference on computer vision and pattern recognition}, 2018, pp. 3550--3558.

\bibitem{chen2020generating}
P.~Chen, Y.~Zhang, M.~Tan, H.~Xiao, D.~Huang, and C.~Gan, ``Generating visually aligned sound from videos,'' \emph{IEEE Transactions on Image Processing}, vol.~29, pp. 8292--8302, 2020.

\bibitem{iashin2021taming}
V.~Iashin and E.~Rahtu, ``Taming visually guided sound generation,'' \emph{arXiv preprint arXiv:2110.08791}, 2021.

\bibitem{luo2023diff}
S.~Luo, C.~Yan, C.~Hu, and H.~Zhao, ``Diff-foley: Synchronized video-to-audio synthesis with latent diffusion models,'' \emph{Advances in Neural Information Processing Systems}, vol.~36, pp. 48\,855--48\,876, 2023.

\bibitem{wang2024frieren}
Y.~Wang, W.~Guo, R.~Huang, J.~Huang, Z.~Wang, F.~You, R.~Li, and Z.~Zhao, ``Frieren: Efficient video-to-audio generation network with rectified flow matching,'' \emph{Advances in Neural Information Processing Systems}, vol.~37, pp. 128\,118--128\,138, 2024.

\bibitem{liu2022flow}
X.~Liu, C.~Gong, and Q.~Liu, ``Flow straight and fast: Learning to generate and transfer data with rectified flow,'' \emph{arXiv preprint arXiv:2209.03003}, 2022.

\bibitem{wang2024v2a}
H.~Wang, J.~Ma, S.~Pascual, R.~Cartwright, and W.~Cai, ``V2a-mapper: A lightweight solution for vision-to-audio generation by connecting foundation models,'' in \emph{Proceedings of the AAAI Conference on Artificial Intelligence}, vol.~38, no.~14, 2024, pp. 15\,492--15\,501.

\bibitem{radford2021learning}
A.~Radford, J.~W. Kim, C.~Hallacy, A.~Ramesh, G.~Goh, S.~Agarwal, G.~Sastry, A.~Askell, P.~Mishkin, J.~Clark \emph{et~al.}, ``Learning transferable visual models from natural language supervision,'' in \emph{International conference on machine learning}.\hskip 1em plus 0.5em minus 0.4em\relax PmLR, 2021, pp. 8748--8763.

\bibitem{elizalde2023clap}
B.~Elizalde, S.~Deshmukh, M.~Al~Ismail, and H.~Wang, ``Clap learning audio concepts from natural language supervision,'' in \emph{ICASSP 2023-2023 IEEE International Conference on Acoustics, Speech and Signal Processing (ICASSP)}.\hskip 1em plus 0.5em minus 0.4em\relax IEEE, 2023, pp. 1--5.

\bibitem{liu2023audioldm}
H.~Liu, Z.~Chen, Y.~Yuan, X.~Mei, X.~Liu, D.~Mandic, W.~Wang, and M.~D. Plumbley, ``Audioldm: Text-to-audio generation with latent diffusion models,'' \emph{arXiv preprint arXiv:2301.12503}, 2023.

\bibitem{cheng2025mmaudio}
H.~K. Cheng, M.~Ishii, A.~Hayakawa, T.~Shibuya, A.~Schwing, and Y.~Mitsufuji, ``Mmaudio: Taming multimodal joint training for high-quality video-to-audio synthesis,'' in \emph{Proceedings of the Computer Vision and Pattern Recognition Conference}, 2025, pp. 28\,901--28\,911.

\bibitem{tian2025audiox}
Z.~Tian, Y.~Jin, Z.~Liu, R.~Yuan, X.~Tan, Q.~Chen, W.~Xue, and Y.~Guo, ``Audiox: Diffusion transformer for anything-to-audio generation,'' \emph{arXiv preprint arXiv:2503.10522}, 2025.

\bibitem{liu2025thinksound}
H.~Liu, J.~Wang, K.~Luo, W.~Wang, Q.~Chen, Z.~Zhao, and W.~Xue, ``Thinksound: Chain-of-thought reasoning in multimodal large language models for audio generation and editing,'' \emph{arXiv preprint arXiv:2506.21448}, 2025.

\bibitem{rayleigh1875our}
L.~Rayleigh, ``On our perception of the direotion of a source of sound,'' \emph{Proceedings of the Musical Association}, vol.~2, no.~1, pp. 75--84, 1875.

\bibitem{leng2022binauralgrad}
Y.~Leng, Z.~Chen, J.~Guo, H.~Liu, J.~Chen, X.~Tan, D.~Mandic, L.~He, X.~Li, T.~Qin \emph{et~al.}, ``Binauralgrad: A two-stage conditional diffusion probabilistic model for binaural audio synthesis,'' \emph{Advances in Neural Information Processing Systems}, vol.~35, pp. 23\,689--23\,700, 2022.

\bibitem{dagli2024see}
R.~Dagli, S.~Prakash, R.~Wu, and H.~Khosravani, ``See-2-sound: Zero-shot spatial environment-to-spatial sound,'' \emph{arXiv preprint arXiv:2406.06612}, 2024.

\bibitem{gao20192}
R.~Gao and K.~Grauman, ``2.5 d visual sound,'' in \emph{Proceedings of the IEEE/CVF Conference on Computer Vision and Pattern Recognition}, 2019, pp. 324--333.

\bibitem{garg2023visually}
R.~Garg, R.~Gao, and K.~Grauman, ``Visually-guided audio spatialization in video with geometry-aware multi-task learning,'' \emph{International Journal of Computer Vision}, vol. 131, no.~10, pp. 2723--2737, 2023.

\bibitem{liu2024visually}
M.~Liu, J.~Wang, X.~Qian, and X.~Xie, ``Visually guided binaural audio generation with cross-modal consistency,'' in \emph{ICASSP 2024-2024 IEEE International Conference on Acoustics, Speech and Signal Processing (ICASSP)}.\hskip 1em plus 0.5em minus 0.4em\relax IEEE, 2024, pp. 7980--7984.

\bibitem{li2024cross}
Z.~Li, B.~Zhao, and Y.~Yuan, ``Cross-modal generative model for visual-guided binaural stereo generation,'' \emph{Knowledge-Based Systems}, vol. 296, p. 111814, 2024.

\bibitem{heydari2024immersediffusion}
M.~Heydari, M.~Souden, B.~Conejo, and J.~Atkins, ``Immersediffusion: A generative spatial audio latent diffusion model,'' \emph{arXiv preprint arXiv:2410.14945}, 2024.

\bibitem{kushwaha2024diff}
S.~S. Kushwaha, J.~Ma, M.~R. Thomas, Y.~Tian, and A.~Bruni, ``Diff-sage: End-to-end spatial audio generation using diffusion models,'' \emph{arXiv preprint arXiv:2410.11299}, 2024.

\bibitem{kimvisage}
J.~Kim, H.~Yun, and G.~Kim, ``Visage: Video-to-spatial audio generation,'' in \emph{ICLR}, 2025.

\bibitem{Liu2025OmniAudio}
H.~Liu, T.~Luo, K.~Luo, Q.~Jiang, P.~Sun, J.~Wang, R.~Huang, Q.~Chen, W.~Wang, X.~Li, S.~Zhang, Z.~Yan, Z.~Zhao, and W.~Xue, ``Omniaudio: Generating spatial audio from 360-degree video,'' \emph{arXiv preprint arXiv:2504.14906}, 2025.

\bibitem{cheng2024yolo}
T.~Cheng, L.~Song, Y.~Ge, W.~Liu, X.~Wang, and Y.~Shan, ``Yolo-world: Real-time open-vocabulary object detection,'' in \emph{Proceedings of the IEEE/CVF Conference on Computer Vision and Pattern Recognition}, 2024, pp. 16\,901--16\,911.

\bibitem{song2025depthmaster}
Z.~Song, Z.~Wang, B.~Li, H.~Zhang, R.~Zhu, L.~Liu, P.-T. Jiang, and T.~Zhang, ``Depthmaster: Taming diffusion models for monocular depth estimation,'' \emph{arXiv preprint arXiv:2501.02576}, 2025.

\bibitem{kong2020diffwave}
Z.~Kong, W.~Ping, J.~Huang, K.~Zhao, and B.~Catanzaro, ``Diffwave: A versatile diffusion model for audio synthesis,'' \emph{arXiv preprint arXiv:2009.09761}, 2020.

\bibitem{chen2020vggsound}
H.~Chen, W.~Xie, A.~Vedaldi, and A.~Zisserman, ``Vggsound: A large-scale audio-visual dataset,'' in \emph{ICASSP 2020-2020 IEEE International Conference on Acoustics, Speech and Signal Processing (ICASSP)}.\hskip 1em plus 0.5em minus 0.4em\relax IEEE, 2020, pp. 721--725.

\bibitem{brinkmann2019hutubs}
F.~Brinkmann, M.~Dinakaran, R.~Pelzer, J.~J. Wohlgemuth, F.~Seipl, and S.~Weinzierl, ``The hutubs hrtf database,'' \emph{DOI}, vol.~10, p. 14279, 2019.

\bibitem{martin2007interpolation}
R.~Martin and K.~McAnally, ``Interpolation of head-related transfer functions,'' Tech. Rep., 2007.

\bibitem{diaz2021gpurir}
D.~Diaz-Guerra, A.~Miguel, and J.~R. Beltran, ``gpurir: A python library for room impulse response simulation with gpu acceleration,'' \emph{Multimedia Tools and Applications}, vol.~80, no.~4, pp. 5653--5671, 2021.

\bibitem{VGGSoundSolo2025}
\BIBentryALTinterwordspacing
L.~Zhao, ``{VGGSound-Solo},'' IEEE DataPort, 2025. [Online]. Available: \url{https://dx.doi.org/10.21227/m57x-cr16}
\BIBentrySTDinterwordspacing

\bibitem{allen1979image}
J.~B. Allen and D.~A. Berkley, ``Image method for efficiently simulating small-room acoustics,'' \emph{The Journal of the Acoustical Society of America}, vol.~65, no.~4, pp. 943--950, 1979.

\bibitem{kong2020panns}
Q.~Kong, Y.~Cao, T.~Iqbal, Y.~Wang, W.~Wang, and M.~D. Plumbley, ``Panns: Large-scale pretrained audio neural networks for audio pattern recognition,'' \emph{IEEE/ACM Transactions on Audio, Speech, and Language Processing}, vol.~28, pp. 2880--2894, 2020.

\bibitem{gemmeke2017audio}
J.~F. Gemmeke, D.~P. Ellis, D.~Freedman, A.~Jansen, W.~Lawrence, R.~C. Moore, M.~Plakal, and M.~Ritter, ``Audio set: An ontology and human-labeled dataset for audio events,'' in \emph{2017 IEEE international conference on acoustics, speech and signal processing (ICASSP)}.\hskip 1em plus 0.5em minus 0.4em\relax IEEE, 2017, pp. 776--780.

\bibitem{koutini2022efficient}
K.~Koutini, J.~Schl{\"u}ter, H.~Eghbal-zadeh, and G.~Widmer, ``Efficient training of audio transformers with patchout,'' \emph{Interspeech 2022}, 2022.

\bibitem{salimans2016improved}
T.~Salimans, I.~Goodfellow, W.~Zaremba, V.~Cheung, A.~Radford, and X.~Chen, ``Improved techniques for training gans,'' \emph{Advances in neural information processing systems}, vol.~29, 2016.

\bibitem{viertola2025temporally}
I.~Viertola, V.~Iashin, and E.~Rahtu, ``Temporally aligned audio for video with autoregression,'' in \emph{ICASSP 2025-2025 IEEE International Conference on Acoustics, Speech and Signal Processing (ICASSP)}.\hskip 1em plus 0.5em minus 0.4em\relax IEEE, 2025, pp. 1--5.

\bibitem{zhao2025dualspec}
L.~Zhao, S.~Chen, L.~Feng, J.~Zhang, X.-L. Zhang, C.~Zhang, and X.~Li, ``Dualspec: Text-to-spatial-audio generation via dual-spectrogram guided diffusion model,'' \emph{arXiv preprint arXiv:2502.18952}, 2025.

\bibitem{lin2014microsoft}
T.-Y. Lin, M.~Maire, S.~Belongie, J.~Hays, P.~Perona, D.~Ramanan, P.~Doll{\'a}r, and C.~L. Zitnick, ``Microsoft coco: Common objects in context,'' in \emph{European conference on computer vision}.\hskip 1em plus 0.5em minus 0.4em\relax Springer, 2014, pp. 740--755.

\bibitem{deng2009imagenet}
J.~Deng, W.~Dong, R.~Socher, L.-J. Li, K.~Li, and L.~Fei-Fei, ``Imagenet: A large-scale hierarchical image database,'' in \emph{2009 IEEE conference on computer vision and pattern recognition}.\hskip 1em plus 0.5em minus 0.4em\relax Ieee, 2009, pp. 248--255.

\bibitem{scheibler2018pyroomacoustics}
R.~Scheibler, E.~Bezzam, and I.~Dokmani{\'c}, ``Pyroomacoustics: A python package for audio room simulation and array processing algorithms,'' in \emph{2018 IEEE international conference on acoustics, speech and signal processing (ICASSP)}.\hskip 1em plus 0.5em minus 0.4em\relax IEEE, 2018, pp. 351--355.

\end{thebibliography}

\end{document}